\documentclass{ws-procs9x6}
\newcommand{\epem}{\ensuremath{e^+e^-}}
\newcommand{\Vub}{\ensuremath{|V_{ub}|}}
\newcommand{\Vcb}{\ensuremath{|V_{cb}|}}

\newcommand{\Vus}{\ensuremath{|V_{us}|}}

\newcommand{\Vts}{\ensuremath{|V_{ts}|}}

\newcommand{\Vtd}{\ensuremath{|V_{td}|}}
\newcommand{\Kz}{\ensuremath{K^0}}
\newcommand{\Kzb}{\ensuremath{\overline{K}{}^0}}
\newcommand{\BzBzb}{\ensuremath{B^0\overline{B}{}^0}}
\newcommand{\Bz}{\ensuremath{B^0}}
\newcommand{\Bzb}{\ensuremath{\overline{B}{}^0}}
\newcommand{\Bd}{\ensuremath{B^0_d}}
\newcommand{\Bs}{\ensuremath{B^0_s}}

\newcommand{\Bp}{\ensuremath{B^+}}
\newcommand{\Bm}{\ensuremath{B^-}}
\newcommand{\qqbar}{\ensuremath{q\overline{q}}}

\newcommand{\nub}{\ensuremath{\overline{\nu}}}

\begin{document}

\title{B Physics and CP Violation}

\author{R.~V. Kowalewski\footnote{\uppercase{W}ork partially
supported by the \uppercase{N}atural \uppercase{S}ciences and
\uppercase{E}ngineering \uppercase{R}esearch
\uppercase{C}ouncil of \uppercase{C}anada.}}

\address{Department of Physics and Astronomy\\
P.O. Box 3055, \\ 
University of Victoria\\
Victoria, BC V8N2X3 CANADA}


%

\maketitle

\abstracts{ These lectures present the phenomenology of $B$ meson
decays and their impact on our understanding of CP violation in the
quark sector, with an emphasis on measurements made at the $\epem$ $B$
factories.  Some of the relevant theoretical ideas such as the
Operator Product Expansion and Heavy Quark Symmetry are introduced,
and applications to the determination of CKM matrix elements given.
The phenomenon of $B$ flavor oscillations is reviewed, and the
mechanisms for and current status of CP violation in the $B$ system is
given.  The status of rare B decays is also discussed.}

\section{Flavor Physics with $B$ Mesons}

$B$ decays provide a sensitive probe of the physics of quark mixing,
described by the unitary Cabibbo-Kobayashi-Maskawa (CKM) mixing matrix
in the Standard Model (SM).  The mixing of the weak and mass eigenstates of
the quarks provides a rich phenomenology and gives a viable
mechanism for the non-conservation of CP symmetry in the decays of
certain hadrons.  
CP asymmetries in $B$ decays can be large and allow a
determination of the magnitude of the irreducible phase in the CKM
matrix.  The pattern of CP asymmetries observed in $B$ decays can be 
compared with the detailed prediction of these asymmetries in the SM 
in an effort to tease out evidence of new physics (NP).
The study of $B$ decays allows direct measurement of
the magnitudes of the elements \Vub\ and \Vcb\ of the CKM matrix.  The
large mass of the top quark allows information on \Vtd\ and \Vts\ to
be extracted via higher order processes like \BzBzb\ oscillations and
decays involving loop diagrams.  These loop decays have accessible
branching fractions (as large as $10^{-4}$) and allow sensitive
searches for physics beyond the SM, as the effect of new
particles or couplings in the internal loops can manifest itself in
modifications to the rates for these decays.  

In addition to probing the flavor sector of the SM, $B$ decays provide
a laboratory for testing our understanding of QCD.  The scale of
the short-distance physics (e.g.~weak $b$ quark decay) is in the
perturbative regime of QCD while the formation of final state hadrons
and the binding of the $b$ quark to the valence anti-quark is clearly
non-perturbative.  Powerful theoretical tools have been developed to
systematically address the disparate scales.  The Operator Product
Expansion (OPE) together with Heavy Quark Symmetry allow perturbative
calculations to be combined with non-perturbative matrix elements, and
provide relationships amongst the non-perturbative matrix elements
contributing to different processes.  This allows some
non-perturbative quantities to be determined experimentally, and leads
to the vibrant interplay between experiment and theory that has
characterized this area of research in recent years.

\section{Quark mixing}

The mixing between the quark mass eigenstates and their weak
interaction eigenstates, parameterized in the unitary $3\times 3$ CKM matrix,
is responsible for flavor oscillations in the neutral $B$ and $K$ mesons
and leads to an irreducible source of CP violation in the SM through
the non-trivial phase of the CKM matrix.  The CKM matrix can be
parameterized by 3 real angles and one imaginary phase.  The presence
of this irreducible phase is an unavoidable consequence of 3
generation mixing.\footnote{The number of irreducible phases for
$N$-generation mixing is $(N-2)(N-1)/2$.}  
Many parameterizations of
the CKM matrix have been suggested.  One choice in widespread use is
in terms of angles $\theta_{12}$, $\theta_{13}$, $\theta_{23}$ and phase
$\delta$ (see [\refcite{PDG}]).  The magnitudes of the elements in the CKM
matrix decrease sharply as one moves away from the diagonal,
suggesting a parameterization\cite{Wolfenstein} in terms of powers of
$\lambda$, the sine of the the Cabibbo angle.  An
improved version\cite{Wolfplusplus} 
of this ``Wolfenstein parameterization''
will be used here.  The starting point is to let
the parameters $\lambda$, $A$, $\rho$ and $\eta$ satisfy the
relations $\lambda=\Vus$, $A \lambda^2=\Vcb$ and $A \lambda^3
(\rho-i\eta)=\Vub$ and to write the remaining elements in terms of
these four parameters, expanded in powers of $\lambda$.  Given that
$\lambda=0.2196\pm 0.0026$\cite{PDG}, 
highly accurate approximate parameterizations
can be obtained keeping only the first one or two terms in the expansion.
The matrix to order $\lambda^5$ is given here.

{\small
\begin{equation}
\left(\begin{array}{ccc}
\rule{0pt}{12pt}1-\frac{1}{2}\lambda^2-\frac{1}{8}\lambda^4 & \lambda & A\lambda^3(\rho-i\eta) \\
\rule{0pt}{12pt}-\lambda+\frac{1}{2}A^2\lambda^5\left[1-2(\rho+i\eta)\right] & 1-\frac{1}{2}\lambda^2-\frac{1}{8}\lambda^4(1+4A^2) & A\lambda^2 \\
\rule{0pt}{12pt}A\lambda^3\left[1-(\rho+i\eta)(1-\frac{1}{2}\lambda^2)\right] & -A\lambda^2+A(\frac{1}{2}-\rho-i\eta)\lambda^4 & 1-\frac{1}{2}A^2\lambda^4 
\end{array}\right)
\end{equation}
}

This parameterization has several nice features.  The smallness of the
off-diagonal elements is taken up by powers of $\lambda$, leaving the
parameters $A$, $\rho$ and $\eta$ of order unity.  It also makes clear
the near equality of the elements $\Vcb$ and $\Vts$, namely
$|V_{ts}/V_{cb}| = 1 + \mathcal{O}(\lambda^2)$.  

The relations dictated by unitarity allow a convenient geometrical 
representation of the CKM parameters.  The product of any row (column) 
of the matrix times the complex 
conjugate of any other row (column) results in three complex
numbers that sum to zero, and can be drawn as a triangle
in the complex plane.  There are three such independent triangles.  Two
of the three have one side much shorter than the others (i.e. have one
side that is proportional to a higher power of $\lambda$ than are the
others), but the remaining triangle, formed by multiplying the first
column by the complex conjugate of the third column, has all sides
of order $\lambda^3$.  It is this triangle that is usually discussed
when considering the impact of experimental measurements on the
parameters of the CKM matrix.  These unitarity relations need to be
verified experimentally; a violation of unitarity would point to new
physics (e.g.~a fourth generation, in which case the $3\times 3$ submatrix
need not be unitary).

\begin{figure}[htb]
\centerline{\includegraphics*{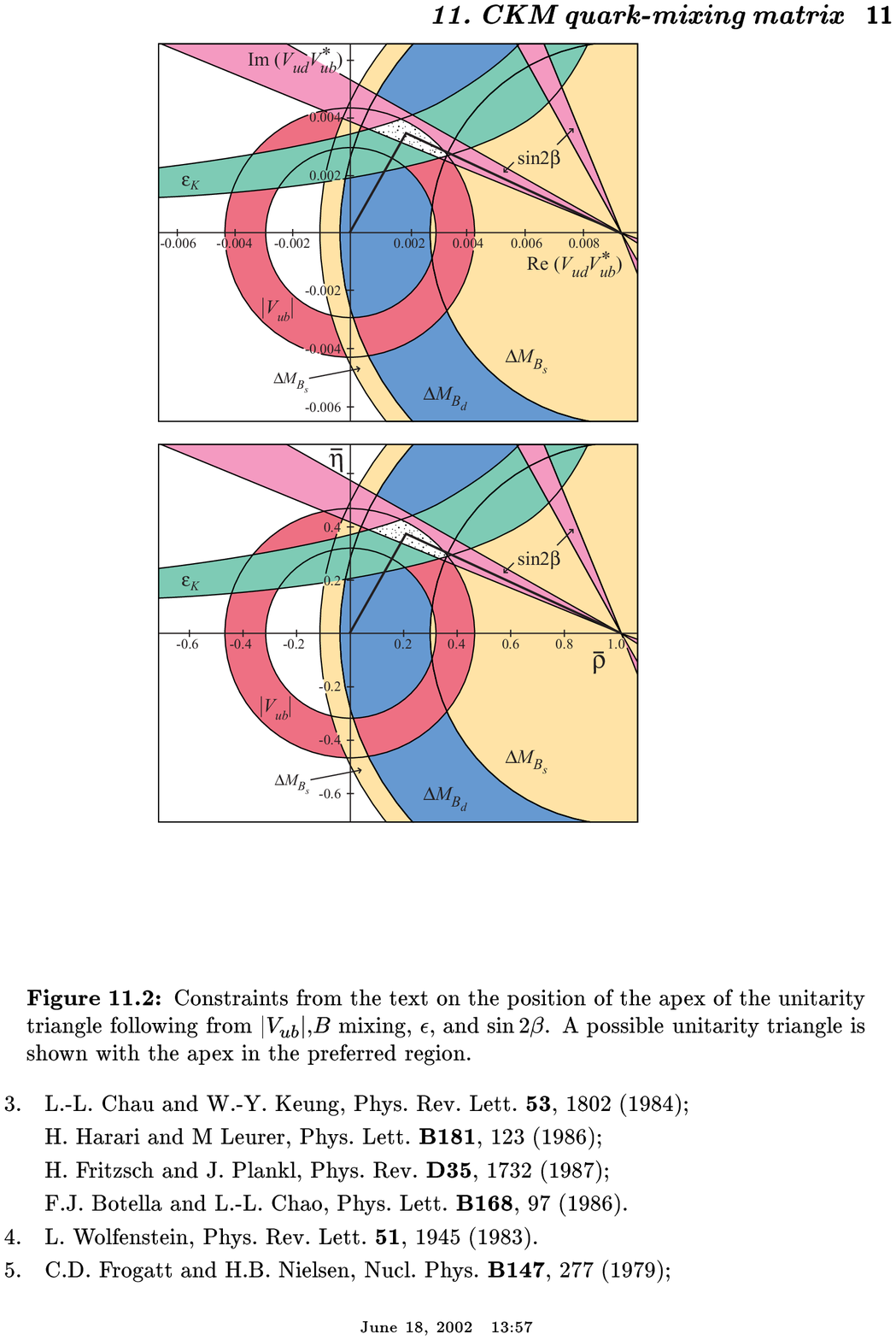}}
\caption{Constraints in the $\overline{\rho}-\overline{\eta}$ plane
from measurements in the $B$ meson and neutral K systems.\label{fig-rhoetaPDG}}
\end{figure}

The unitarity triangle of interest, namely
$V_{ud}V_{ub}^*+V_{cd}V_{cb}^*+V_{td}V_{tb}^*=0$,
is usually rescaled by dividing through by $V_{cd}V_{cb}^*$, giving
a triangle whose base has unit length and lies along the $x$
axis and whose apex is at the point $\frac{V_{ud}V_{ub}^*}{V_{cd}V_{cb}^*}
=1+\frac{V_{td}V_{tb}^*}{V_{cd}V_{cb}^*}$.  Casting these relations
in terms of the parameters $\lambda, A, \rho$ and $\eta$ and defining
$\overline{\rho}=(1-\lambda^2/2)\rho$ and
$\overline{\eta}=(1-\lambda^2/2)\eta$, the apex of the triangle is
at $(\overline{\rho},\overline{\eta})$ up to
corrections of order $\lambda^2$.  By dividing out $V_{cb}$ we largely
eliminate dependence on the parameter $A$, which is in any case
relatively well known\cite{PDG} ($A=0.85\pm 0.04$).
The sides and angles of the unitarity triangle can be expressed
as
\begin{eqnarray}
R_u &=& \frac{V_{ud}V_{ub}^*}{V_{cd}V_{cb}^*} \simeq -\sqrt{\overline{\rho}^2+\overline{\eta}^2} e^{i\gamma} \\
R_t &=& \frac{V_{td}V_{tb}^*}{V_{cd}V_{cb}^*} \simeq -\sqrt{(1-\overline{\rho})^2+\overline{\eta}^2} e^{i\beta} \\
\gamma&=& \mathrm{arg}V_{ub}^* \\
\beta&=& \mathrm{arg}V_{td} \\
\alpha&=&\pi-\beta-\gamma 
\end{eqnarray}
The angles $\alpha, \beta$ and $\gamma$ are also 
known as $\phi_2, \phi_1$ and $\phi_3$, respectively.
Figure~\ref{fig-rhoetaPDG} shows the constraints on the unitarity
triangle given in Ref.~[\refcite{PDG}].

Measurements of $\Vub/\Vcb$ constrain the length of $R_u$ and
constrain $(\overline{\rho},\overline{\eta})$ to lie in an annulus
centered on the origin.  These measurements are improving as more data
and additional theoretical insights are brought to bear.  The bands
emanating from the point (1,0) correspond to the measurement of
$\sin 2\beta$, the result of the impressive initial success of the
$B$-factory program.  These measurements already give some of the most
precise information about the unitarity triangle and are perhaps an
order of magnitude away from being limited by systematics.  The $B$
system also allows access to the length of $R_t$ through measurements
of $\BzBzb$ oscillations, which give indirect sensitivity to $\Vtd$
and $\Vts$.  The determination of $R_t$ is dominated by theoretical
uncertainties.  Measurements of $\Bs$ oscillations and of the ratio of
$b\to d\gamma$ to $b\to s\gamma$ decays should allow improved
determinations of $R_t$.  While $\Bs$ oscillations are not accessible
at $\epem$ $B$ factories, there are good prospects for this measurement
at high energy hadron collider experiments.  Measurement of
the CKM-suppressed $b\to d\gamma$ decays will require significantly
larger data sets than are currently available.

\section{B factory basics}

The $\epem$ $B$ factories operate at the $\Upsilon(4S)$ resonance, a
quasi-bound $b\overline{b}$ state just above the threshold for
production of $\Bp\Bm$ or $\Bz\Bzb$ pairs.  The width of the
$\Upsilon(4S)$ is comparable to the beam energy spread of $\epem$
colliders, making the peak cross-section a weak function of the
machine parameters.  The cross-section at the peak is about 1.1~nb,
to be compared with the underlying \qqbar ($q=u,d,s,c$) cross-section
of $\sim 3.5$~nb.
The $\Upsilon(4S)$ is expected to decay with a branching fraction of
0.5 to each of $\Bp\Bm$ and $\Bz\Bzb$; measurements are consistent
with this expectation, but have a large uncertainty\cite{PDG} 
($\pm14\%$ on the
ratio of the two branching fractions).  The mass difference
between the $\Upsilon(4S)$ and a pair of $B$ mesons is about 20~MeV, so
no additional particles (apart from low-energy radiated photons) are
produced, and the final state $B$ mesons have one unit of orbital
angular momentum and small velocity ($\beta\simeq 0.06$) in the
$\Upsilon(4S)$ rest frame.

The cross-section for $b$ hadron production at hadron colliders is
much higher---typically by a factor of 100 or more, depending on
the collider energy and the acceptance in pseudo-rapidity.  In
addition, $\Bs$ and $b$ baryons are also produced.  However, the
fraction of events containing $b$ hadrons is much lower (a few per mil)
and only those events leaving a clean trigger signature are accessible.
In the short term the most important $B$ physics 
measurements that will be made at the hadron colliders are the
$\Bs$ oscillation frequency and CP asymmetries in $\Bs$ decay.

\subsection{Asymmetric \epem\ $B$ factories}

The asymmetric $B$ factories designed to study CP asymmetries in $B$
decays collide low energy positiron beams (3.1-3.5~GeV) with higher
energy electron beams (9.0-8.0~GeV) and give the center-of-mass of the
collision a boost of $\beta\gamma\simeq 0.5$ along the beam axis.  As
a result the distance between the decay points of the two $B$ mesons,
whose lifetimes are $\sim 1.6$~ps, is typically $250~\mu$m.  This is
critical in measuring CP asymmetries that arise in the interference
between amplitudes for $\BzBzb$ mixing and $B$ decay, since the time
integral of these CP asymmetries vanishes.  The required luminosity is
set by the scale of the branching fractions for $\Bz$ decay to
experimentally accessible CP eigenstates and the need to determine the
flavor of the other $B$ in the event and the time difference between
the two $B$ decays.  This leads to a requirement of 30~fb$^{-1}$/year,
or a luminosity of at 
least $\mathcal{L}=3\times 10^{33}$~cm$^{-2}$~s$^{-1}$.  The
$B$ factory 
detectors must provide good particle identification over the full momentum
range, excellent vertex resolution, and good efficiency and resolution
for charged tracks and photons.

Two $B$ factories exceeding these minimum requirements have been operating
since 1999: the Belle detector at the KEK-B accelerator at KEK and the
BaBar detector at the PEP-II accelerator at SLAC.  Both facilities
have been operating well, providing unprecedented luminosities (as of
February, 2003):\\
\begin{center}
\begin{tabular}{lcc}
 & KEK-B / Belle & PEP-II / BaBar \\
$\mathcal{L}_\mathrm{max}\ (10^{33}\mathrm{cm^{-2}s^{-1}})$ & 8.3 & 4.8 \\
best day (pb$^{-1}$) & 434 & 303 \\
total (fb$^{-1}$) & 106 & 96 
\end{tabular}
\end{center}

The BaBar and Belle detectors are broadly similar: each has a silicon
vertex detector surrounded by a wire tracking chamber with a
Helium-based gas, particle identification, a CsI(Tl) calorimeter, a
superconducting coil and detectors interspersed with the iron flux
return to measure $K^0_L$ and identify muons.  They differ in their
silicon detectors, where Belle uses 3 layers at small radii (3.0-5.8~cm)
while BaBar adds two layers at large radii (up to 12.7 and 14.6~cm);
in each case the wire tracking device starts a few centimeters beyond the last
silicon layer.  The biggest difference is in the technology used for
particle identification.  While both detectors exploit dE/dx in the
tracking chambers to identify low momentum hadrons, Belle uses a combination
of time-of-flight and aerogel threshold Cherenkov counters.  
BaBar has a different strategy,
using long quartz bars to generate Cherenkov light from passing
particles and to transport it via total internal reflection to a
water-filled torus where the Cherenkov angle is measured.
The performance of each detector is similar in most respects.

\section{B hadron decay}

The weak decay of a free $b$ quark is completely analogous to muon decay.
While there are no free quarks, this is a useful starting point for
understanding $B$ hadron decay.  The $B$ meson lifetime is relatively large 
($\sim 1.6$~ps), much longer than the lighter $D$ mesons or tau lepton,
due to the smallness of \Vcb.  Semileptonic decays are prominent, 
with branching fractions of
about 10.5\%\ each for semi-electronic and semi-muonic decays and about
2.5\%\ for semi-tauonic decays.  These decays contain a single hadronic
current and are more tractable theoretically than fully hadronic
final states, which make up the bulk of the $B$ decay rate.  Purely
leptonic decays are helicity suppressed (less so for tau modes) and
require either CKM-suppressed $b\to u$ transitions or Flavor-Changing
Neutral-Currents (FCNC), 
forbidden at tree level in the SM, leading to branching
fractions\cite{leptonicBF} 
$\mathcal{B}(\Bp\to\tau^+\nu_\tau)\sim 6\times 10^{-5}$,
$\mathcal{B}(\Bz\to\tau^+\tau^-)\sim 3\times 10^{-8}$.

The long $B$ lifetime, along with the large top quark mass which breaks
the GIM mechanism, give radiative ``penguin'' FCNC decays
like $b\to s\gamma$ and $b\to s\ell\overline{\ell}$ branching fractions
in the $10^{-4}-10^{-6}$ range, making them accessible.  These same
factors result in large $\BzBzb$ mixing through second order
weak processes involving box diagrams with virtual $W$ and top particles,
and make rare $B$ decays a fruitful ground for searching for physics
beyond the SM.

\subsection{Theoretical picture of $B$ decay}

The material in this subsection is covered in more detail in several
excellent review articles.\cite{HQETreviews}
The free quark picture is a useful starting point, but the impact of
the strong interactions binding the $b$ quark in the $B$ hadron must be
addressed to achieve a quantitative understanding of $B$ decay.  Early
attempts to do this involved models that gave the $b$ quark an r.m.s.
``Fermi momentum'' to account for bound state effects.  While these
models improved the understanding of some observables, like the
lepton energy spectrum in semileptonic decay, they did not provide a
quantitative assessment of the theoretical uncertainties.  Since
the early 1990s new theoretical methods have been used based on
a systematic separation of short-distance and long-distance scales
through the Operator Product Expansion (OPE).  The OPE formalism is used
to develop effective field theories where the short distance behaviour
is integrated out of the theory.  
A scale $\mu$ is defined to separate
the long- and short-distance regimes.  The choice of $\mu$ is in 
principle arbitrary, since no observables can depend on it {\em if
the calculation is carried out to all orders.}  In practice, $\mu$ is
chosen to satisfy $\Lambda_{\mathrm{QCD}}\ll \mu \ll M_W$.
The result of integrating out the short-distance heavy particle fields
is a non-local action that is then
expanded in a series of local operators of increasing dimension
whose (Wilson) coefficients contain the short-distance physics.
Perturbative corrections (e.g.~for hard gluon emission) 
to the short-distance physics are incorporated
using renormalization-group methods.\cite{RenormGroup} 
$B$ decay amplitudes are expressed as
\begin{equation}
A(B\to F) = \langle F|H_{\mathrm{eff}}|B\rangle
= \sum_i C_i(\mu)\langle F|Q_i(\mu)|B\rangle
\end{equation}
The Wilson coefficients $C_i(\mu)$
typically include leading-log or next-to-leading-log
corrections.  The sum involves increasing powers of the heavy
quark mass, required to offset the increasing dimensions of the
non-perturbative operators.  This suppression of higher-dimension
operators is central to our ability to make quantitative predictions
for $B$ decays.  At present only terms of order $1/m_b^2$ or $1/m_b^3$
nnare considered, leaving only a modest number
of non-perturbative matrix elements to determine (either experimentally
or through non-perturbative calculational techniques).

A major step forward in the understanding of $B$ decays was the
recognition of heavy quark symmetry.  The scale 
$\Lambda_{\mathrm{QCD}}\sim 0.2$~GeV at which QCD becomes
non-perturbative is small compared to the heavy quark mass $m_Q$.  
As a result
the gluons binding the heavy quark and light spectator are too soft to
probe the quantum numbers---mass, spin, flavor---of the heavy quark.
In the limit $m_Q\to \infty$ the heavy quark degrees of freedom decouple
completely from the light degrees of freedom, resulting in a spin-flavor
SU(2$N_h$) symmetry, where $N_h$ is the number of heavy quark flavors\footnote{
Effectively two, since the top quark decays before hadronizing.}  
This symmetry has an analogue in atomic physics, where the nucleus
acts as a static source of electric charge and where nuclear
properties decouple from the degrees of freedom associated with the
electrons; to first approximation different isotopes or nuclear spin
states of an element
have the same chemistry.  In heavy-light systems, the heavy quark
acts as a static source of color charge.

Heavy quark symmetry forms the basis of an effective field theory of
QCD, namely Heavy Quark Effective Theory (HQET).  The key observation is that
in the heavy quark limit the velocities of the heavy quark and the
hadron containing it are equal: $p_Q=m_Qv+k$, where $p_Q$ is the 4-vector
of the heavy quark, $v$ is the 4-velocity of the hadron and $k$ is the
residual momentum, whose components are small compared to $m_Q$.  The
degrees of freedom associated with energetic ($\mathcal{O}(2m_Q)$)
fluctuations of the heavy quark field are integrated out, resulting
in an effective Lagrangian
\begin{equation}
L_{\mathrm{eff}}=\overline{h}_viv\cdot Dh_v+\frac{1}{2m_Q}\overline{h}_v 
(i{\not\!\!D}_\perp)^2 h_v + \frac{g_s}{4m_Q}\overline{h}_v\sigma_{\mu\nu}
G^{\mu\nu} h_v + \mathcal{O}(m_Q^{-2})
\end{equation}
where $h_v(x)=e^{im_Qv\cdot x}\frac{1+\not v}{2}Q(x)$ is the upper 2
components of the heavy quark Dirac spinor, the lower components having
been integrated out of the theory.  The first term is all that remains
in the limit $m_Q\to\infty$, and is manifestly invariant under SU(2$N_h$).
The second term is the kinetic energy operator $\boldmath{O}_K$
for the residual motion
of the heavy quark, and the third term gives the operator $\boldmath{O}_G$
for the interaction of the heavy
quark spin with the color-magnetic field.  The matrix elements associated
with these operators are non-perturbative, but can be related to measurable
quantities.

The non-perturbative parameters at lowest order are $\lambda_1 = 
\langle Q|\boldmath{O}_K|Q\rangle /2m_Q$ and
$\lambda_2 = -\langle Q|\boldmath{O}_G|Q\rangle /6m_Q$.  The
mass of a heavy meson can be written
\begin{equation}
m_{H_Q} = m_Q + \overline{\Lambda} + 
\frac{-\lambda_1+2\left[J(J+1)-\frac{3}{2}\right]\lambda_2}{2m_Q}
+\mathcal{O}\left(m_Q^{-2}\right)
\end{equation}
The parameter $\overline{\Lambda}$ arises from the light quark degrees
of freedom, and is defined by 
$\overline{\Lambda}=\lim_{m_Q\to\infty}(m_{H_Q}-m_Q)$.  In the
heavy quark limit all systems with the same light quark degrees of
freedom should have the same $\overline{\Lambda}$.  It can be verified
that this gives a good description of SU(3)$_\mathrm{flavor}$
breaking ($m(B_s)-m(B_d)\simeq m(D_s)-m(D_d)$).  The mass splitting
between the vector and pseudo-scalar mesons determines $\lambda_2$ to
be approximately 0.12~GeV:
\begin{eqnarray}
m^2(B^*)-m^2(B) &=& 0.49\mathrm{~GeV^2} = 4\lambda_2+\mathcal{O}(m_b^{-1})\\
m^2(D^*)-m^2(D) &=& 0.55\mathrm{~GeV^2} = 4\lambda_2+\mathcal{O}(m_c^{-1})
\end{eqnarray}

\subsection{Exclusive semileptonic decays and \Vcb}

Heavy quark effective theory is clearly a powerful tool in
understanding transitions between two heavy-light systems, since the
light degrees of freedom don't see the change in heavy quark flavor or
spin in the heavy quark limit.  An area of great practical importance
is the determination of $\Vcb$ from exclusive semileptonic decays like
$B\to \overline{D}^*\ell^+\nu$.  
In the heavy quark limit the form factor for the
decay can depend only on the product of 4-velocities of the initial
and final mesons, $w=v_B\cdot v_{D^*}$.  This universal form factor is
known as the Isgur-Wise function\cite{IsgurWise} $\xi(w)$.  The form
of the function is not specified in HQET, but its normalization is unity
at the ``zero-recoil'' point, $\xi(1)=1$, where the $D^*$ meson is
stationary in the $B$ meson rest frame and the light degrees of
freedom are blind to the change in heavy quark properties.

One of the striking predictions of HQET is that the four independent
form factors in a general $P\to V\ell\nub$ transition are all related
to the Isgur-Wise function:
\begin{eqnarray*}
h_V(w) &=& h_{A_1}(w) = h_{A_3}(w) = \xi(w)\\
h_{A_2}(w)&=&\left[\frac{2m_Bm_{D^*}(1+w)}{(m_B+m_{D^*})^2}\right]\xi(w)\to 0
\mathrm{\ as\ }m_Q\to \infty
\end{eqnarray*}
where $h_V$ is the form factor for the vector current and $h_{A_1}$, 
$h_{A_2}$ and $h_{A_3}$ are the form factors associated with the
axial-vector current.  These relations can be tested experimentally.

The normalization of the physical form factor at zero recoil differs from
unity due to QCD radiative corrections and heavy quark symmetry-breaking
corrections.  This leads to
\begin{eqnarray}
\mathcal{F}(w) &=& \xi(w)+\mathcal{O}(\alpha_s(m_Q))+\mathcal{O}\left((\Lambda_{\mathrm{QCD}}/m_Q)^2\right)+...\\
\mathcal{F}(1) &=& \eta_A\left(1+C\frac{\Lambda^2_{\mathrm{QCD}}}{m_Q^2}\right)
\end{eqnarray}
where $\eta_A$ encorporates the QCD correction\cite{Czarnecki} and 
the absence of a correction at order $\Lambda_{\mathrm{QCD}}/m_Q$ is known
as Luke's theorem.\cite{LukeTheorem}  The value $\mathcal{F}(1)$ must
be calculated using non-perturbative techniques such as Lattice QCD
or QCD sum rules.  These lead to the currently accepted 
value\cite{IsgurWiseNorm} of
$0.91\pm 0.04$.  This rather precise prediction can be combined with
an experimental measurement of the decay rate for 
$B\to \overline{D}^*\ell^+\nu$
to determine $\Vcb^2$.  The measurement is complicated by the need to
extrapolate the differential decay rate $d\Gamma/dw$ 
to the point $w=1$, since $d\Gamma/dw$ vanishes there.  A further
complication comes from the fact that the transition pion from the
$D^*\to D$ decay has very low momentum in the $B$ frame for $w\simeq 1$.
Nevertheless, the measured experimental rates give 
$\mathcal{F}(1)\Vcb=(38.3\pm1.0)\times 10^{-3}$ from which a precise
value of \Vcb\ is obtained:\cite{PDG}
\begin{equation}
\Vcb = (42.1\pm 1.1\pm 1.9)\times 10^{-3}
\end{equation}

Similar measurements can be made for $B\to \overline{D}\ell^+\nu$.  
The theoretical
situation here is less favorable, since Luke's theorem does not prevent
$\Lambda_{\mathrm{QCD}}/m_Q$ corrections in this case, and the experimental
situation is complicated by feed-down from 
$B\to \overline{D}^*\ell^+\nu$ decays.
The ratio of form factors at zero recoil can be measured and compared
with predictions:
\begin{eqnarray*}
\frac{\mathcal{G}(1)}{\mathcal{F}(1)} &=& 1.08\pm 0.06\ (\mathrm{theory})\\
&=& 1.08\pm 0.09\ (\mathrm{experiment})
\end{eqnarray*}
where $\mathcal{G}(w)$ is the form factor in $B\to \overline{D}\ell^+\nu$ 
decay.
The tests of the HQET predictions for form factors in semileptonic
$B$ decays to charm are nearly at the point of testing 
the symmetry breaking terms, and are improving.

\subsection{Inclusive semileptonic decays}

The OPE and HQET formalism can be used to study inclusive semileptonic
decays.  Bound state effects can be accounted for in a systematic
expansion in terms of $\alpha_S$ and $1/m_b$.  To do this, however,
one must introduce a new assumption, namely that the (long distance)
process of forming color singlet final state hadrons does not change
the rates calculated at the quark level.  This assumption goes under
the name of quark-hadron duality.  While it has been demonstrated to
hold under certain conditions --- e.g.~in the cross-section for
$\epem\to$hadrons and in tau decays to hadrons --- it also clearly
breaks down in regions where the density of color-singlet final states
is small or zero, e.g.~for $m_{\mathrm{had}} < 2m_\pi$.  While duality
violations in inclusive rates are expected to be small, their level is
hard to quantify, and they can be important when a severely restricted
phase space is examined (e.g.~when selecting $b\to
u\ell\nub$ decays by requiring that the charged lepton momentum
exceeds the kinematic endpoint for leptons from $b\to c\ell\nub$
decays).

The semileptonic decay rate in the Heavy Quark Expansion (HQE) is
\begin{equation}
\Gamma(B\to X\ell^+\nu)=\frac{G_F^2m_b^5}{192\pi^3}\left[
1+C_1\frac{\alpha_s(m_b)}{\pi}+...+\frac{-\lambda_1-9\lambda_2}{2m_b^2}+...
\right]
\end{equation}
where the non-perturbative matrix elements $\lambda_1$ and $\lambda_2$
are familiar from HQET.  Note the absence
of a $1/m_b$ correction term; this allows the term in brackets to be computed
to a precision of about 5\%.  The dependence on $m_b^5$ can result in
large uncertainties in the theoretical prediction.  These have been brought
under control by using the upsilon expansion,\cite{upsilonexpansion} in
which (ignoring the subtleties) one-half the
mass of the $\Upsilon(1S)$ is substituted for $m_b$.  The resulting
theoretical expression for extracting $\Vub$ from the corresponding inclusive
semileptonic decay width is\cite{VubToBF}
\begin{equation}
\Vub = (3.87\pm 0.10\pm 0.10)\times 10^{-3}
\times\sqrt\frac{\Gamma(\overline{B}\to X_u\ell\nub)}{1.0\mathrm{ns}^{-1}}
\end{equation}
A similar expression relates \Vcb\ to $\Gamma(\overline{B}\to X_c\ell\nub)$:
\begin{equation}
\Vcb = (44.5\pm 0.8\pm 0.7\pm 0.5)\times 10^{-3}
\times\sqrt\frac{\Gamma(\overline{B}\to X_c\ell\nub)}{65.6\mathrm{ns}^{-1}}
\end{equation}

Measuring the semileptonic width for $b\to c\ell\nu$ transitions
is straight-forward; up to small corrections one just measures the
semileptonic branching fractions and lifetimes of $B$ mesons and
sets $\Gamma_{\mathrm{sl}}=\mathcal{B}_{\mathrm{sl}}/\tau_B$.
The momentum spectrum of leptons from $B$ decay is fairly stiff, while
leptons from other processes (most notably $b\to c\to \ell$ decays)
are softer.
The formula given above can then be used to extract $\Vcb$ with
small theoretical uncertainty.  Proceeding in this manner with the
present experimental information\cite{PDG} on $\tau(\Bz)=1542\pm 16$~fs,
$\tau(\Bp)=1674\pm 18$~fs and 
$\mathcal{B}(\overline{B}\to X_c\ell\nub)=(10.38\pm 0.32)\%$ gives\cite{PDG}
\begin{equation}
\Vcb = (40.4\pm 0.5_{\mathrm{exp}}\pm 0.5_{\mathrm{non-pert}}\pm
0.8_{\mathrm{pert}})\times 10^{-3}
\end{equation}
This determination is fully consistent with the value derived from
exclusive semileptonic decays.

The determination of \Vub\ from inclusive semileptonic decays
is more challenging due to the
large background from the CKM-favored $b\to c$ transitions.  The first
measurements of \Vub\ came from the endpoint of the lepton momentum
spectrum, where a small fraction ($\sim 10\%$) of the leptons from
$b\to u$ transitions lie above the endpoint for leptons from $b\to c$
transitions.  The experimental signal in this region is very robust, but
the limited acceptance results in large theoretical uncertainties
in extracting \Vub.  These uncertainties arise because of limited knowledge
of the so-called shape function, i.e. the distribution of the (virtual)
$b$ quark mass in the $B$ meson, which affects the kinematic distributions
of the final state particles and therefore changes the fraction $f_u$ of
$b\to u$ decays above the minimum accepted lepton momentum.

One means of reducing the theoretical uncertainty is to obtain
information on the shape function from other $B$ decays.  The easiest
method conceptually is to examine the photon energy spectrum in the
$B$ rest frame for $b\to s\gamma$ decays.\cite{btosgammaenergy} 
Since the photon does not
undergo strong interactactions it probes the $b$ quark properties,
with $m_b = 2E_\gamma$ at lowest order.  The first moment of this
spectrum is essentially the mean $b$ quark mass (or, equivalently,
$\overline{\Lambda}$) while the second moment is essentially
$-\lambda_1$.  Similar use can be made\cite{q2El} of the recoiling $W$ in
semileptonic $b\to u$ decays, where $m_b = E_W+|\vec{p}_W|$, but
requires the reconstruction of the neutrino momentum.  
Measurements of other moments in semileptonic decays of both $b\to u$
and $b\to c$ (e.g.~of $E_\ell$, $m^2_{\mathrm{had}}$...)
also give information on the non-perturbative parameters $\overline{\Lambda}$
and $\lambda_1$ and can be used to constrain the range of variation that
must be considered in assessing the theoretical error due to the
acceptance cuts.

Another approach to reducing the theoretical uncertainty in \Vub\ from
inclusive semileptonic decays is to measure more than just the charged
lepton.  Setting aside the experimental difficulties, a measurement of
the mass of the recoiling hadron allows a much greater fraction of the
$b\to u$ final states to be accepted, namely those with invariant mass
$m_{\mathrm{had}}<m_D$, thereby reducing theoretical
uncertainties.\cite{mhadVub}  Measuring the invariant mass of the lepton and
neutrino ($q^2$) and requiring $q^2>(m_B-m_D)^2$, while having a lower
acceptance than a cut on $m_{\mathrm{had}}$, results in a similar
theoretical uncertainty.\cite{q2El}  
Both of these approaches are in progress
at the $B$ factories.  These approaches should yield $\Vub$ with
uncertainties of 10\%\ or less in the near future.

Exclusive charmless semileptonic decays ($B\to\pi\ell^+\nu$, etc.)
provide an independent means of determining \Vub.  The experimental
measurements of these decays are improving.  At present the leading
uncertainties come from theoretical calculations and models of
form factors.  There is an expectation that lattice QCD calculations
of these form factors will eventually allow \Vub\ to be extracted
with uncertainties of less than 10\%, providing an independent test
of the \Vub\ determined from inclusive semileptonic decays.

\section{\BzBzb\ oscillations}

The material developed in the next two sections is covered in greater
depth in several excellent
reviews.\cite{CPMixingReviews,BurasErice,BurasReview}  Quark flavor is
not conserved in weak interactions.  As a result, transitions are
possible between neutral mesons and their antiparticles.  These
transitions result in the decay eigenstates of the
particle--anti-particle system being distinct from their mass
eigenstates.  In systems where the weak decay of the mesons is
suppressed (e.g.~by small CKM elements) and the $\Delta$(flavor)$=2$
transitions between particle and anti-particle are enhanced (due to
the large top mass and favorable CKM elements) the decay eigenstates
can be dramatically different from the mass eigenstates, resulting in
the spectacular phenomenon of flavor oscillations.  The flavor
oscillations first observed in the neutral $K$ system result in the
striking lifetime difference between the two decay eigenstates and the
phenomenon of regeneration.  In neutral $B$ mesons, the lifetime
differences are small as is the branching fraction to eigenstates of
CP.  As a result, oscillations are observed by studying the
time evolution of the flavor composition ($b$ or $\overline{b}$) of
weak decays to flavor eigenstates.

The neutral $B$ mesons form a 2-state system, with the flavor eigenstates
denoted by
\begin{equation}
|\Bz\rangle = \left(\begin{array}{c}1\\ 0\end{array}\right)\quad\quad\quad
|\Bzb\rangle = \left(\begin{array}{c}0\\ 1\end{array}\right)
\end{equation}
The effective Hamiltonian, which includes the $2^{\mathrm{nd}}$-order
weak $\Delta b=2$ transition, is diagonalized in the mass eigenbasis,
obtained by solving
\begin{equation}
H|B_{\mathrm{H,L}}\rangle = E_{\mathrm{H,L}}|B_{\mathrm{H.L}}\rangle
\end{equation}
where the subscript H and L denote the ``heavy'' and ``light'' eigenstates
and the effective Hamiltonian can be written as
\begin{eqnarray}
H &=& 
\left(\begin{array}{cc}M_{11} & M_{12}\\ M_{12}^* & M_{22}\end{array}\right)
-\frac{i}{2}  \nonumber
\left(\begin{array}{cc}\Gamma_{11} & \Gamma_{12}\\ \Gamma_{12}^* & \Gamma_{22}\end{array}\right)\\
&=& \left({M}-\frac{i}{2}{\Gamma}\right)
\left(\begin{array}{cc}1&0\\ 0&1\end{array}\right) +
\left(\begin{array}{cc}0&M_{12}-\frac{i}{2}\Gamma_{12}\\
                       M_{12}^*-\frac{i}{2}\Gamma_{12}^* &0\end{array}\right)
\end{eqnarray}
where in the last line CPT symmetry is used to write $M=M_{11}=M_{22}$
and $\Gamma=\Gamma_{11}=\Gamma_{22}$.  The values of $M$ and $\Gamma$
are determined by the quark masses and the strong and electromagnetic
interactions.  The last term induces $\Delta b=2$ transitions and is
responsible for flavor oscillations.  The dispersive ($M_{12}$) and
absorptive ($\Gamma_{12}$) parts correspond to virtual and real
intermediate states, respectively.  The time evolution of a state that
is a pure $\Bz$ at $t=0$ is given by

{\small
\begin{eqnarray}
|\Bz (t)\rangle &=& e^{-(\Gamma /2-iM)t} \left\{  \nonumber
\cos\frac{\Delta Mt}{2} \cosh\frac{\Delta\Gamma t}{4}
-i\sin\frac{\Delta Mt}{2}\sinh\frac{\Delta\Gamma t}{4}\right\} |\Bz\rangle\\
+&\frac{q}{p}& e^{-(\Gamma/2-iM)t} \left\{
i\sin\frac{\Delta Mt}{2} \cosh\frac{\Delta\Gamma t}{4}
-\cos\frac{\Delta Mt}{2}\sinh\frac{\Delta\Gamma t}{4}\right\} |\Bzb\rangle
\label{eq:bzoft}
\end{eqnarray}}

\noindent where 
(ignoring CP violation for the moment) $|p|^2+|q|^2=1$,
$M=\frac{1}{2}(M_{\mathrm{H}}+M_{\mathrm{L}})$,
$\Gamma=\frac{1}{2}(\Gamma_{\mathrm{H}}+\Gamma_{\mathrm{L}})$,
$\Delta m=\frac{1}{2}(M_{\mathrm{H}}-M_{\mathrm{L}})$, and
$\Delta\Gamma=\frac{1}{2}(\Gamma_{\mathrm{H}}+\Gamma_{\mathrm{L}})$.
In the $B$ system we always have $\Delta\Gamma \ll \Gamma$, since
the branching fraction to flavor-neutral intermediate states (with
quark content $c\overline{c}d\overline{d}$ or 
$u\overline{u}d\overline{d}$) is $\mathcal{O}(1\%)$, and only these
can contribute to $\Delta\Gamma$.  
The formula above simplifies in this approximation.
No such constraint applies to
$\Delta m/\Gamma$, since virtual intermediate states contribute.  In
fact, the large top quark mass breaks the GIM mechanism that would
otherwise cancel this FCNC and enhances $\Delta m$.  

The dominant diagrams responsible for $\BzBzb$ oscillations in the
SM are $W$-$t$ box diagrams.  While the short-distance physics in
these diagrams can be calculated perturbatively, there are also
non-perturbative matrix elements that enter the width.  The standard
OPE expression for $\Delta m$ is\cite{BurasReview}
\begin{equation}
\Delta m_q = \frac{G_F^2}{6\pi^2} \eta_B (f^2_{B_q}\hat{B}_{B_q}) M_W^2
S_0(m_t^2/M_W^2) \left|V_{tq}\right|^2
\end{equation}
where $\eta_B$ is a perturbative QCD correction, 
$f_B$ is the $B$ meson decay constant, 
$\hat{B}_B$ is the ``bag factor'', and $q$ can be either $d$ or $s$.
The term $S_0$ is a known function of $m_t^2/M_W^2$ that has a
value of $\sim 2.5$ for the measured masses.  The uncertainty in
$f_B^2\hat{B}_B$ produces a $\sim 30\%$ uncertainty in extracting
$\Vtd$ or $\Vts$ by comparing $\Delta m_q$ with measured values.

Many methods have been used to study $\BzBzb$ oscillations.  The most
sensitive is the dilepton charge asymmetry\cite{Delta_md} as a
function of the $B^0$ decay time (or, in the case of the $B$
factories, the decay time difference between the two $B$ mesons).  The
current world average\cite{PDG} for the oscillation frequency is
$\Delta m_d = (0.489\pm 0.008)\,\mathrm{ps}^{-1}$.  Improvements in
calculations of $f_B^2\hat{B}_B$ are needed in order to improve the
impact this measurement has on constraining the unitarity triangle.

$\Bs$ oscillations cannot be studied in $\Upsilon(4S)$ decays.
The $0^{\mathrm{th}}$-order expectation for $\Delta m_s$ is 
$\Delta m_s = |V_{ts}/V_{td}|^2 \Delta m_d \sim 15\,\mathrm{ps}^{-1}$,
but the numerical value is not very precise due to uncertainties in the
CKM elements.  The large value of $\Delta m_s$ implies very rapid oscillations
and makes the measurement of $\Delta m_s$ challenging.  Note that
evidence exists for $\Bs$ flavor change;\cite{Bsmixing} it is consistent
with being maximal.  However, at present there are only
lower limits\cite{PDG} 
on $\Delta m_s$ from experiments at LEP and from CDF.  Similar uncertainties
in $f_B^2\hat{B}_B$ arise when comparing the theoretical and measured values
of $\Delta m_s$ as for $\Delta m_d$.  However, the theoretical uncertainty
on the ratio $\Delta m_d/\Delta m_s$ is smaller, as some of the uncertainties
in the ratio $(f_{B_s}^2\hat{B}_{B_s})/(f_{B_d}^2\hat{B}_{B_d})$ 
cancel; this is
an active area of theoretical investigation.\cite{FBratio}
A measurement of $\Delta m_s$ will provide significant information
in constraining the unitarity triangle.

\section{CP violation}

The CP operation takes particle into anti-particle, admitting an
arbitrary (and unobservable) phase change:
\begin{equation}
\mathrm{CP}|X\rangle  = e^{+2\theta_{\mathrm{CP}}}|
\overline{X}\rangle;\ \ \ \mathrm{CP}
|\overline{X}\rangle = e^{-2\theta_{\mathrm{CP}}}|X \rangle
\end{equation}
CP violation was discovered\cite{Cronin} in $K^0_L$ decays to $\pi\pi$
final states in 1964.  At the time there was no known mechanism
that could accommodate this observation.  Such a mechanism was
introduced by Kobayashi and Maskawa\cite{KM} in 1973.  They postulated
the existence of a third generation of quarks, noting that the
$3\times 3$ unitary matrix describing the mixing of quark weak and
mass eigenstates retains one phase that cannot be removed by rephasing
the quark fields.  The presence of this phase allows for CP to be
violated in reactions involving two or more interfering
amplitudes.
The first particle of the third generation was discovered the very
next year\cite{Perl} and the first quark of the third
generation\cite{Lederman} three years thereafter.  The mechanism
uncovered by Kobayashi and Maskawa in fact {\em requires} CP violation in
the absence of special values of the non-trivial phase or other
fine-tuning conditions, like quark mass degeneracies or the vanishing
of a CKM angle.  

CP violation has been observed more recently in the
$B$ system\cite{firstCPB,BaBarsin2b,Bellesin2b}, and appears to
conform to the expectations of the KM picture.  It
remains to be determined, however, if the KM mechanism is the
sole source of the CP violation seen in the $K$ and $B$ systems.  

An invariant measure of the size of the CP violation in the CKM
matrix is given by the Jarlskog invariant\cite{Jarlskog}
\begin{equation}
J=c_{12}c_{23}c^2_{13}s_{12}s_{23}s_{13}\sin\delta\simeq A^2\lambda^6\eta
\end{equation}
where $c_{ij}$ and $s_{ij}$ are shorthand for the cosine and sine of
the angle $\theta_{ij}$, and $A$, $\lambda$ and $\eta$ are the parameters
of the Wolfenstein parameterization.  The maximum value of $J$ in any
unitary $3\times 3$ matrix is $(6\sqrt{3})^{-1}\sim 0.1$; the value in the
CKM matrix is $\sim 4\times 10^{-5}$, which underlies the statement
that CP violation in the SM is small.  

\subsection{CP violation in $B$ decay}

The CP violation in the SM is the result of a phase, and is therefore
only observable in processes involving interfering amplitudes.  The
mechanisms for generating this interference in $B$ decays fall into
three classes: CP violation in flavor mixing, CP violation in interfering
decay amplitudes, and CP violation in the interference between flavor
mixing and decay.  

The CP violation first observed in the neutral $K$ system arises in
flavor mixing.  The mass eigenstates in the $K$ system are not 
quite eigenstates
of CP, and can be described as having a small component with the opposite
CP.  This type of CP violation arises because of the interference between
the on-shell and off-shell $\Delta s=2$ transitions between $\Kz$
and $\Kzb$.  In the $B$ system the 
smallness of $\Delta\Gamma / \Gamma$ (i.e. the on-shell
transition width) suppresses this source of CP
violation (in constrast to the neutral $K$ system, where 
$\Delta\Gamma_K\simeq -2\,\Delta m_K$).  

CP will be violated in $\Bz\Bzb$ mixing only to the extent
that $|q/p|\ne 1$ (see eq.~\ref{eq:bzoft}):
\begin{eqnarray}
\left|\frac{q}{p}\right|^2=
\frac{\langle\Bzb|H_{eff}|\Bz\rangle}{\langle\Bz|H_{eff}|\Bzb\rangle}=
\left|\frac{M_{12}^*-\frac{i}{2}\Gamma_{12}^*}
{M_{12}-\frac{i}{2}\Gamma_{12}}\right|\ne 1
\end{eqnarray}
A manifestation of this CP violation would be an asymmetry in the
semileptonic decays,
\begin{equation}
A_{\mathrm{CP}}=\frac{\Gamma(\Bzb\to\ell^+\nu X)-\Gamma(\Bz\to\ell^-\overline{\nu} X)}{
\Gamma(\Bzb\to\ell^+\nu X)+\Gamma(\Bz\to\ell^-\overline{\nu} X)}
\end{equation}
which is proportional to $(1-|q/p|^4)/(1+|q/p|^4)=\Im(\Gamma_{12}/M_{12})$.
In the SM this asymmetry
is $\mathcal{O}(10^{-4})$ and cannot be calculated with precision due to
large hadronic uncertainties in the determination of $\Gamma_{12}$.

Interference between competing decay amplitudes can also render
CP violation observable in the SM.
This is known as direct CP violation, and was first seen\cite{epsilonprime}
as a non-zero
value of $\epsilon'/\epsilon$ 
in neutral $K$ decays.  In direct CP violation
$|\overline{A}_{\overline{f}}/A_f|\ne 1$, implying
the decay rate for
$\overline{B}\to\overline{f}$ is not the same as for $B\to f$.
All unstable particles (not just neutral mesons) 
admit this type of CP violation in principle.
The decay rate asymmetry is
\begin{equation}
A_{\mathrm{CP}}=\frac{\Gamma(B\to f)-\Gamma(\overline{B}\to\overline{f})
}{\Gamma(B\to f)+\Gamma(\overline{B}\to\overline{f})}\propto
2|A_1||A_2|\sin\delta\sin\phi
\end{equation}
where $A_1$ and $A_2$ are competing decay amplitudes that lead
to the same final state $f$, $\delta$ is the difference between the
strong interaction phases of the two amplitudes, and $\phi$ is the
difference between the weak phases (which include the CKM contribution).
Once again, precise predictions of this type of CP violation are
not available due to hadronic uncertainties in calculating the
strong phase difference $\delta$.  It remains of interest, however,
to search for decay modes where the weak phase difference $\phi\simeq 0$,
in which case any $A_{\mathrm{CP}}$ which deviates significantly from zero
is a sign of new physics.

The CP violation that arises due to the interference between
$\BzBzb$ mixing and decay is different, in that theoretical
predictions nearly free from strong interaction uncertainties
are possible.  The interference arises between the
decay $\Bz\to f_{\mathrm{CP}}$ and the sequence $\Bz\to\Bzb\to f_{\mathrm{CP}}$.
Clearly only those final states that are eigenstates of CP can contribute.
This type of CP violation can be present even when $|q/p|=1$
and $\overline{A}_{\overline{f}_{\mathrm{CP}}}/A_{f_{\mathrm{CP}}}=1$.  The quantity
\begin{equation}
\lambda_{f{_{\mathrm{CP}}}}=\eta_{f_{\mathrm{CP}}}\frac{q}{p}\frac{\overline{A}_{\overline{f}_{\mathrm{CP}}}}
{A_{f_{\mathrm{CP}}}}
\end{equation}
where $\eta_{f_{\mathrm{CP}}}=\pm 1$ is the CP eigenvalue,
is independent of phase conventions and contains the information
on CP violation.  Since the interference is mediated by mixing,
the CP asymmetry has a characteristic time dependence:
\begin{eqnarray}
A_{\mathrm{CP}}(t)&=&\frac{\Gamma(\Bzb(t)\to f_{\mathrm{CP}})-\Gamma(\Bz(t)\to f_{\mathrm{CP}})}{
\Gamma(\Bzb(t)\to f_{\mathrm{CP}})+\Gamma(\Bz(t)\to f_{\mathrm{CP}})} \nonumber\\
&=& \frac{1-|\lambda_{f_{\mathrm{CP}}}|^2}{1+|\lambda_{f_{\mathrm{CP}}}|^2}
\cos\Delta m_B t
+ \frac{2\Im\lambda_{f_{\mathrm{CP}}}}{1+|\lambda_{f_{\mathrm{CP}}}|^2}
\sin\Delta m_B t
\label{eq:ACPt}
\end{eqnarray}
The coefficient of the $\cos\Delta m_B t$ term is a measure of
direct CP violation (or CP violation in mixing, but this is always
small in the $B$ system).  For decay modes
dominated by a single decay amplitude it vanishes.  
The coefficient
of the $\sin\Delta m_B t$ term is then a pure phase.  This phase
can be related to angles in the unitarity triangle with very little
theoretical uncertainty.  This is why the study of CP violation in
neutral $B$ decays has attracted so much attention.  
It's worth seeing how this comes about.

The ratio $q/p$ is given by $\sqrt{M_{12}^*/M_{12}}$ (ignoring the
small contribution of $\Gamma_{12}$), where 
$M_{12}\propto V_{tb}^*V_{td}e^{i(\pi-2\theta_{\mathrm{CP}})}$, so
$q/p = e^{2i(\theta_{\mathrm{CP}}+\theta_M)}$ is a pure phase.  $\theta_M$
is the phase difference coming from CKM elements and $\theta_{\mathrm{CP}}$
is the arbitrary phase change in the CP transformation.  The 
decay amplitudes are 
\begin{eqnarray}
A_{f_{\mathrm{CP}}}&=&\langle f_{\mathrm{CP}}|H|\Bz\rangle =|A|e^{i(\delta+\phi_D)}\\
\overline{A}_{f_{\mathrm{CP}}}&=&\langle f_{\mathrm{CP}}|H|\Bzb\rangle =
\eta_{\mathrm{CP}}e^{-2i\theta_{\mathrm{CP}}}|A|e^{i(\delta-\phi_D)}
\end{eqnarray}
giving 
$\overline{A}_{f_{\mathrm{CP}}}/A_{f_{\mathrm{CP}}}=\eta_{\mathrm{CP}}e^{-2i(\theta_{\mathrm{CP}}+\theta_D)}$,
where $\theta_D$ is the weak phase associated with the decay.  Note
that the dependence on the strong phase $\delta$ cancels in the
ratio of decay amplitudes.
Putting this together,
$\lambda_{f_{\mathrm{CP}}}=\eta_{\mathrm{CP}}e^{2i(\theta_M-\theta_D)}$.  This clean
relation holds for $\Bz$ decays to CP eigenstates that proceed via
a single decay amplitude.  Channels involving interfering decay
amplitudes in general result in a mix of direct CP violation and
CP violation in the interefence of mixing and decay, and in these
cases the correspondence with unitarity triangle angles is not free
from hadronic uncertainties.

\subsection{Measuring Unitarity Angles with CP Asymmetries}
\subsubsection{$\sin 2\beta$}

The ``golden mode'' for studying CP violation in $B$ decays is
$\Bz\to J/\psi K^0_S$.  In this case the decay is dominated by the
tree diagram with internal $W\to c\overline{s}$ emission, leading to 
$(\overline{b}d)\to \overline{c}W^+d\to (\overline{c}c)(\overline{s}d)$.
This is not a flavor-neutral state at the quark level,
but becomes so at the hadron level through $K^0$ mixing.  
For this decay one finds
\begin{eqnarray}
\lambda_{f_{\mathrm{CP}}} &=& \eta_{\mathrm{CP}}
\left(\frac{V_{tb}^*V_{td}}{V_{tb}V_{td}^*}\right)_{\mathrm{B\ mixing}}
\left(\frac{V_{cs}^*V_{cb}}{V_{cs}V_{cb}^*}\right)_{\mathrm{decay}}
\left(\frac{V_{cd}^*V_{cs}}{V_{cd}V_{cs}^*}\right)_{\mathrm{K\ mixing}}\\
\Im(\lambda_{f_{\mathrm{CP}}})&=& \eta_{\mathrm{CP}}\sin 2\beta
\end{eqnarray}
with very little theoretical uncertainty.  The CP eigenvalue is -1
(for $\Bz\to J/\psi K^0_L$ it is +1).  This decay mode is also
favorable experimentally.  The product branching fraction
$\mathcal{B}(\Bz\to J/\psi
K^0_S)\mathcal{B}(J/\psi\to\ell^+\ell^-)\simeq 5\times 10^{-5}$ is
well within the reach of $B$ factories, and the final state includes a
lepton pair, enabling excellent background suppression.  

The experimental determination of $\sin 2\beta$ involves three key
elements:
\begin{itemize}
\item The reconstruction of the CP eigenstate, $J/\psi K^0_S$.  This
requires good momentum and energy resolution for charged particles
and photons.
\item The determination of the $b$ quark flavor of the recoiling
$B$ meson at the time of its decay.  This requires good particle
identification in order to cleanly identify the charged kaons and
leptons that are used to infer the $b$ quark flavor.
\item The determination of the difference between the decay times of
the $B$ decay to the CP eigenstate and the recoiling $B$.  This
requires an asymmetric collider to boost the pair of $B$ mesons
along the beam direction and excellent vertex resolution to extract
the spatial distance between the decay points of the two $B$ mesons.
\end{itemize}
The $B$ factories have been optimized to make this measurement.  Given
a sample of $B$ decays to CP eigenstates, a determination of the flavor
of the recoiling $B$, and the time difference between the decays, one
can form the CP asymmetry 
\begin{eqnarray}
A_{\mathrm{CP}}(\Delta t)&=&\frac{N(\Bz)-N(\Bzb)}{N(\Bz)+N(\Bzb)}\nonumber \\
&\simeq& (1-2w)\, \sin 2\beta\, \sin\Delta m_B \Delta t\ \ .
\end{eqnarray}
The asymmetry is based on the flavor assignment ($\Bz$ or
$\Bzb$) of the recoiling $B$, $w$ is the probability of incorrectly
determinating the flavor of the recoiling $B$, and $\Delta t$ is the
time difference between the two $B$ decays.  One must determine the
dilution $(1-2w)$ in order to extract $\sin 2\beta$.  Since the
underlying asymmetry is an oscillatory function, its amplitude is also
diminished by the imperfect resolution on the decay time difference.
Both of these effects (flavor mis-tagging and $\Delta t$ resolution)
can be controlled by considering a related asymmetry formed using
fully reconstructed $\Bz$ decays to flavor eigenstates (like
$\Bz\to D^{*-}\pi^+$) in place of the CP eigenstate sample:
\begin{eqnarray}
A_{\mathrm{mixing}}(\Delta t)&=&\frac{N(\mathrm{mixed})-N(\mathrm{unmixed})}
{N(\mathrm{mixed})+N(\mathrm{unmixed})} \nonumber \\
&\simeq& (1-2w)\, \cos\Delta m_B \Delta t
\end{eqnarray}
The flavor eigenstate sample has much higher statistics than the
CP eigenstate sample, ensuring precise determinations of $w$ and
the $\Delta t$ resolution.  In practice the dilution factor at
the $B$ factories is $(1-2w)\sim 0.25$.

\begin{figure}[htb]
\centerline{
\includegraphics*[scale=0.7]{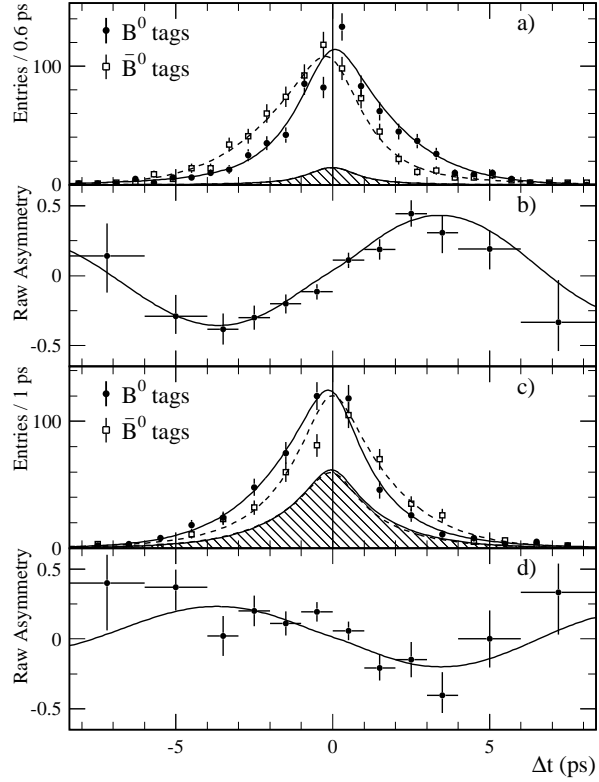}}
\caption{The distributions of $\Delta t$ for events with recoiling $B$
mesons tagged as $\Bz$ and $\Bzb$ (a) and the CP asymmetry (b) for
$J/\psi K^0_S$ decays.  The same distributions are shown in (c) and (d)
for $J/\psi K^0_L$ decays.  
This figure is taken from Ref.~[\protect\refcite{BaBarsin2b}].
\label{fig-CPasymmetry}}
\end{figure}

The BaBar and Belle experiments first observed non-zero 
CP violation\cite{firstCPB} in 2001 and have now measured
$\sin 2\beta$ ($\sin \phi_1$) with good precision.\cite{BaBarsin2b,Bellesin2b}
\begin{eqnarray}
\sin 2\beta &=& 0.719\pm 0.074\pm 0.035\ \ \mathrm{(Belle)}\\
\sin 2\beta &=& 0.741\pm 0.067\pm 0.034\ \ \mathrm{(BaBar)}
\end{eqnarray}
The systematic errors are dominated by the statistics of the auxiliary
samples used to evaluate them, and should continue to fall with increasing 
luminosity.  Figure~\ref{fig-CPasymmetry} shows the 
CP asymmetry observed in BaBar for both the $J/\psi K^0_S$ and
$J/\psi K^0_L$ modes.  The $K^0_L$ mode is reconstructed by using the
position of the $K^0_L$ interaction in the calorimeter or muon
system and the known $B$ energy to estimate the $K^0_L$ 4-vector and
form the invariant mass of the
$J/\psi K^0_L$ pair.  While the background is higher than for the
$J/\psi K^0_S$ sample, it's satisfying to see a CP asymmetry of
the opposite sign!

The measurements of $\sin 2\beta$ now give precise constraints
in the $\overline{\rho}-\overline{\eta}$ plane (see 
Figure~\ref{fig-rhoetaPDG}).

\subsubsection{$\sin 2\alpha$}

There is substantial effort at present on determination of the angle
$\alpha$ ($\phi_2$).  This angle is accessible in $b\to u$ transitions
leading to $u\overline{u}d\overline{d}$ final states, e.g.~$\Bz\to
\pi^+\pi^-$.  In contrast to the golden mode, however, there are both
tree and penguin decay amplitudes that contribute appreciably to these
decays, rendering the precise determination of $\alpha$ more
difficult.  The experimental situation is also less favorable 
for several reasons:
the very small branching fractions\cite{BelleBtopipi,BaBarsin2a}
($\mathcal{B}(\Bz\to\pi^+\pi^-)\simeq 5\times 10^{-6}$); the
higher backgrounds, primarily from the underlying $e^+e^-\to
q\overline{q}$ continuum events; and the difficulty in distinguishing
$B^0\to\pi^+\pi^-$ decays from the more numerous $B^0\to K^+\pi^-$ decays.
The amplitudes of the underlying penguin and tree diagrams can be
constrained\cite{isospinpipi} by measuring the branching fractions of
the isospin-related channels $\Bz\to\pi^+\pi^-$, $\Bz\to\pi^0\pi^0$
and $B^+\to\pi^+\pi^0$.  This is, however, challenging; at
present\cite{Btopizpiz} only upper limits exist on
$\mathcal{B}(\Bz\to\pi^0\pi^0)$.  Other approaches\cite{QuinnSnyder}
to limiting the uncertainty due to the penguin amplitudes have been
proposed; this is an active area of investigation.  In practice one
measures the coefficients $C_{\pi\pi}$ and $S_{\pi\pi}$ of the cosine
and sine terms\footnote{ This is another unfortunate case of
notational differences.  BaBar and Belle use different conventions
(related to the inclusion or not of the CP eigenvalue in the
definition of the coefficients) resulting in
$C_{\pi\pi}^{\mathrm{BaBar}} = -A_{\pi\pi}^{\mathrm{Belle}}$.}  in
eq.~\ref{eq:ACPt}.  The coefficienct of the
sine term would be $\sin 2\alpha$ in the
absence of penguin contributions.

\begin{figure}[htb]
\centerline{
\includegraphics*[scale=0.8,trim=0 330 0 0]{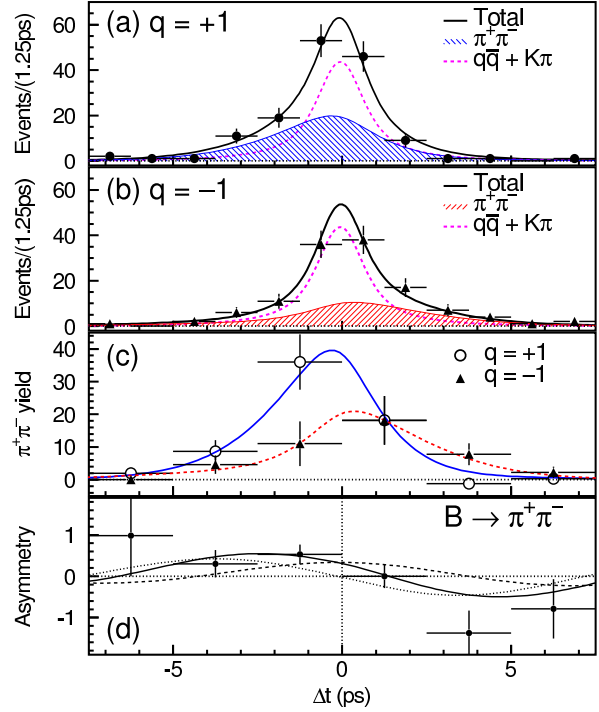}}
\caption{Distributions of $\Bz\to\pi^+\pi^-$ events versus $\Delta t$
for $\Bz$ tags (a) and $\Bzb$ tags (b).  The background-subtracted
distributions and the asymmetry are shown in (c) and (d).  
This figure is taken from Ref.~[\protect\refcite{Bellesin2a}].
\label{fig-BelleCPpipi}}
\end{figure}

The current measurements of these coefficients in BaBar and Belle
differ by about $2.5\sigma$, and lead to rather different interpretations
on the evidence for direct CP violation in $\Bz\to\pi^+\pi^-$ 
decays:\cite{BaBarsin2a,Bellesin2a}
\begin{eqnarray}
S_{\pi\pi} &=&+0.02\pm 0.34\pm 0.05\ \ \mathrm{(BaBar)}\\
S_{\pi\pi} &=&-1.23\pm 0.41\ ^{+0.08}_{-0.07}\ \ \ \ \mathrm{(Belle)}\\
C_{\pi\pi} &=&-0.30\pm 0.25\pm 0.05\ \ \mathrm{(BaBar)}\\
-A_{\pi\pi}&=&-0.77\pm 0.27\pm 0.08\ \ \mathrm{(Belle)}
\end{eqnarray}
Note that the bound $S^2_{\pi\pi}+C^2_{\pi\pi}$ must be satisfied
by the true values of these coefficients.  The Belle data are
shown in Figure~\ref{fig-BelleCPpipi}.

\subsection{CP Asymmetries in $\Bs$ Decays}

The $\Bs$ system can also be used to study CP violation.   
However, $\Bs$ production is suppressed relative to $\Bd$, implying smaller
signals and higher backgrounds.  The outlook for studying $\Bs$ at threshold
$\epem$ machines is not good.  However, dedicated experiments at high energy
hadron colliders are expected to contribute significantly in this area.

The presence of spectator $s$ quark makes a different set of 
unitarity angles accessible in $\Bs$ decays.  The rapid oscillation 
term ($\Delta m_{\Bs}\sim 30\Delta m_{B^0_d}$) makes time resolved 
experiments difficult, but not impossible.
The width difference between the $\Bs$ mass eigenstates
may be exploited as well.

\section{Rare decays}

Rare $B$ decays offer a window on new physics.  A particularly fruitful
place to look for new physics is in FCNC decays, which are highly suppressed
in the Standard Model.  New physics can enhance these rates substantially;
see Refs.~[\refcite{rareDecays}] and~[\refcite{BurasErice}]
for overviews of this subject.
The first $b\to s$ FCNC decays observed,\cite{btosgamma}
namely $b\to s\gamma$, still
provide the best limit on the mass of a charged Higgs boson.  The
penguin decays $b\to s\gamma$, $b\to s\ell^+\ell^-$ ($\ell=e$ or $\mu$) and
$b\to s\nu\overline{\nu}$ are areas of active investigation at the $B$
factories.  The CKM-suppressed modes, with $s$ replaced by $d$, have not
yet been observed; their measurement will allow clean determinations of
the ratio $\Vtd^2/\Vts^2$.

The branching fraction for $b\to s\gamma$ is now measured\cite{btosgamma} and
predicted\cite{btosgammatheory} with good precision:
\begin{eqnarray}
\mathcal{B}(b\to s\gamma) &=& (3.3\pm 0.4)\times 10^{-4}\ \ \mathrm{experiment}\\
                          &=& (3.29\pm 0.33)\times 10^{-4}\ \ \mathrm{theory}
\end{eqnarray}
The $B$ factories have recently observed the first evidence for FCNC
decays involving $Z$ penguin and $W$ box diagrams.  Belle has measured\cite{Belle-bsll}
the inclusive decay $\mathcal{B}(b\to s\ell^+\ell^-)=(6.1\pm
1.4\,^{+1.4}_{-1.1})\times 10^{-4}$ 
and the combined BaBar\cite{BaBar-BKll} and Belle\cite{Belle-BKll}
measurements give $\mathcal{B}(B\to K\ell^+\ell^-)=(7.6\pm 1.8)\times 10^{-7}$.
These are compatible with SM expectations.\cite{rareDecays}
The sensitivity will improve
with increasing luminosity and the forward-backward asymmetry of the
lepton pairs will be measured, providing stringent constraints on new
physics.  

The related decay $b\to s\nu\overline{\nu}$ is also of interest.  It is
very clean theoretically, and is sensitive to all three generations.
The SM rate is also higher than for $b\to s\ell^+\ell^-$ due to the
larger couplings of neutrinos to the $Z$.  The experimental signature,
which includes at least two missing particles, makes searches for these
modes also sensitive to exotic, non-interacting particles (e.g.~the
lightest supersymmetric particle).  The best existing limit is on the
mode $B^+\to K^+\nu\overline{\nu}$, where 
BaBar has a preliminary result\cite{Knunu}
of $\mathcal{B}(B^+\to K^+\nu\overline{\nu})<9.4\times 10^{-5}$ at 90\%\ c.l.  
The sensitivity
of these searches is improving as data sets increase.

\section{Summary}

The $B$ factory program has had a very fruitful beginning---and it
is still the beginning.  The next few years will bring significant
advances in our knowledge of flavor physics.

CP asymmetries in $B$ decays have been observed, are large and will be
observed in many modes in the coming years.  Precision studies of $B$
decays and oscillations provide the most significant source of
information on three of the four CKM parameters and are beginning to provide
stringent tests of the Standard Model.  The interplay of theory and
experiment is vibrant and necessary in order to extract precision
information on the matrix elements $\Vub$ and $\Vcb$ and on the angles
$\alpha$ and $\gamma$ of the unitarity triangle.  Rare $B$ decays
offer a good window on new physics due to the large top
quark mass, and the sensitivity
to these decays is improving rapidly.  $B$ hadrons are also a
laboratory for studying QCD at large and small scales, and a
theoretical framework has been developed to make precise predictions
and suggest new measurements to test the soundness of the framework.

The constraints of time have dictated an abbreviated treatment
of a number of the topics covered here and have precluded the
inclusion of others.  The space devoted to the wealth of $B$
physics measurements made at non-$B$ factory facilities has
been minimal; I can only refer the interested reader to broader
reviews of $B$ physics, e.g. as given in Ref.~[\refcite{CKMworkshop}].

The field of flavor physics is vibrant at the start of the
$21^{\mathrm{st}}$ century.  The $B$ factories and neutrino
experiments, in discovering CP violation in $B$ decays and neutrino
oscillations, have produced the most significant discoveries since the
LEP/SLC program.  These same two fields will probe deeper into the
mysteries of flavor oscillations and CP violation throughout this
decade.  Flavor physics is becoming precision physics!

\section*{Acknowledgements}

I'd like to thank the organizers of the Lake Louise Winter Institute for
their kind invitation and hospitality.

\end{document}